\documentclass[aps,onecolumn,a4paper,oneside,preprint,tightenlines,
nobibnotes,nofootinbib,amsfonts,amssymb,amsmath,eqsecnum,showpacs,
showkeys,epsfig]{revtex4}
\usepackage{graphicx}
\usepackage{hyperref}
\newcommand{\ef}{\operatorname{Erf}}
\begin{document}
\title{ Nonsingular potentials from excited state factorization of a quantum system with position dependent mass}
\author{Bikashkali Midya}
\email{bikash.midya@gmail.com}
\affiliation{Physics and Applied Mathematics Unit, Indian Statistical
 Institute,\\ Kalkata 700108, India.}
 \begin{abstract}
  The modified factorization technique of a quantum system characterized by position-dependent mass Hamiltonian is presented. It has been shown that the singular superpotential defined in terms of a mass function and a excited state wave function of a given position-dependent mass Hamiltonian can be used to construct non-singular isospectral Hamiltonians. The method has been illustrated with the help of a few examples.
\end{abstract}
 \maketitle
\section{introduction}
The one dimensional time independent position dependent mass Schr\"{o}dinger equation (PDMSE) \cite{BD66}
\begin{equation}
H\psi(x) = \left[\frac{1}{2 m_0}\left(\hat{p}\frac{1}{ m(x)} \hat{p}\right) + V(x)\right] \psi = E \psi,
\end{equation}
[$\hat{p} = -i\hbar \frac{d}{dx}$ being the momentum operator and $m(x)$, the dimensionless form of mass function $M(x) = m_0 m(x)$],
has attracted a lot of
attention during the past few years due to its applications in describing the dynamics of electrons in
many condensed-matter systems \cite{GK93}. The concept of position dependent mass comes from the effective mass approximation \cite{Pr65} which is an useful tool for studying the motion of carrier electrons in pure crystals and also for
the virtual-crystal approximation in the treatment of homogeneous alloys as well as in graded mixed semiconductors. This field has also arisen due to the impressive development in crystallographic growth techniques \cite{GG04}  which allow the production of non uniform semiconductor specimen with abrupt heterojunctions (see e.g., \cite{Ba88} for a review). In different areas of possible applications of low dimensional structures as already
mentioned, there is need to have Hamiltonians with predetermined energy spectrum. For example, in the quantum well
profile optimization \cite{Ra01}, isospectral potentials are generated through supersymmetric quantum mechanics. This is necessary because intersubband optical transitions in a quantum well may be grossly enhanced by
achieving the resonance conditions e.g. appropriate spacings between the most relevant states and also by
tailoring the wave functions so that the matrix elements relevant for this particular effect
are maximized \cite{Bo90}. Consequently, there has been
a growing interest in obtaining exactly solvable PDM Hamiltonians through the usage of supersymmetric quantum mechanics (SUSYQM) \cite{Pl99,MI99,Go02,GN07}, point canonical transformation \cite{Al02,Ba05,Qu09}, Lie algebra \cite{RR02}, Heun equation \cite{DC99} and intertwining method \cite{MR10,SH08} (and references cited therein). However, despite being studied widely, the number of exactly
solvable PDM Hamiltonian is still rather few.

Fairly recently, the standard factorization technique \cite{IH51,An84,FG04} has been modified \cite{BU10} so that the excited state wave function of a given nonsingular constant mass Hamiltonian can be used to construct nonsingular isospectral partner Hamiltonian. In general the superpotential defined by excited state wave function gives rise to the singular partner Hamiltonians with negative energy eigenvalues and usual degeneracy of the energy spectrum is lost \cite{JR84,Ca95}. In ref.\cite{BU10}, the nonuniqueness of the factorization of a constant mass Hamiltonian is used to make it possible to construct acceptable(nonsingular) isospectral Hamiltonians. This algorithm is independent of the choice of wavefunctions used to define the superpotential. Since exactly solvable PDM Hamiltonians are few in number, it should be interesting therefore, to extend the modified factorization technique to handle cases with position-dependent
mass. It may be noted that
because of the presence of a non-constant mass the algorithm in the case of PDMSE is different
from the constant mass case and here our objective is to
examine whether or not factorization based on excited states can be used to search for physically acceptable Hamiltonian within the context
of PDMSE.

In order to make this report self contained, in section IIA we have briefly recalled the notion of usual factorization technique of a quantum system with position dependent mass. In section IIB, we have generalized the usual factorization technique so that excited state wave functions can be used to construct new exactly solvable PDM Hamiltonians. Furthermore, as
an useful application of the proposal, in Section III, we have considered two examples. Section IV is kept for summary. Throughout this report we shall consider $\hbar = 2 m_0 =1.$

\section{Factorization in quantum system with Position Dependent Mass}
\subsection{Usual factorization}
\noindent
 The minimal version of PDM SUSYQM is based on ground state wave function and the charge operators are represented by $Q = A_0^+ \sigma_-$ and $Q^\dag = A_0^- \sigma_+$, where $\sigma_{\pm}= \sigma_1 \pm i \sigma_2$ are the combinations of pauli matrices $\sigma_{1,2}$ and the associated two first order linear operators $A_0^\pm$ are given by \cite{Pl99}
\begin{equation}
A^\pm_0 = \pm\frac{1}{\sqrt{m}}\frac{d}{dx} - \left(\frac{1 \mp 1}{2}\right) \left(\frac{1}{\sqrt{m}}\right)' + W_0 .\label{e8}
\end{equation}
where `prime' denotes the differentiation with respect to $x$. The components of the supersymmetric Hamiltonian ${\bf{H}} = diag(H^-,H^+)$ can be factorized in terms of $A_0^+$ and $A_0^-$ to obtain
\begin{equation}
\displaystyle H_0^\mp = A_0^\mp A_0^\pm = -\frac{1}{m}\frac{d^2}{dx^2} + \frac{m'}{m^2} \frac{d}{dx}+ V_0^\mp (x)
\end{equation}
where without loss of generality we have considered zero ground state energy of $H_0^-$ and the two partner potentials $V_0^\mp$ can be written in terms of superpotential $W_0$ and mass function, as
\begin{equation}
\displaystyle V_0^\mp (x) = W_0^2 \mp \left(\frac{W_0}{\sqrt{m}}\right)' + \left (\frac{1 \mp 1}{2}\right) \left[ \frac{W_0 m'}{m^{\frac{3}{2}}}- \frac{1}{\sqrt{m}}\left(\frac{1}{\sqrt{m}}\right)''\right].\label{e1}
\end{equation}
Clearly, $V_0^+ = V_0^- + \frac{2 W_0'}{\sqrt{m}} - \frac{1}{\sqrt{m}}\left(\frac{1}{\sqrt{m}}\right)''$. The ground state wave function $\psi_0^{-}$ of the Hamiltonian $H_0^-$ determines the superpotential
\begin{equation}
W_0(x) = - \frac{\psi_0^{{-'}}}{\sqrt{m} \psi_0^{-}}~,\label{e6}
\end{equation}
which implies that for a physically acceptable mass function\footnote{The mass function should be positive and without any singularity through out its domain of definition.} the superpotential and hence two partner potentials $V_0^\mp$ are free from singularity (provided the given potential $V_0^-$ is without any singularity). It is well established that when SUSY is unbroken then the spectrum of the two PDM partner Hamiltonians $H^\mp$ are degenerate except for the lowest energy of $H_0^-$. Moreover, the energy eigenvalues and normalized eigenfunctions of the SUSY partner Hamiltonians $H_0^\pm$ are related by \cite{Pl99}
\begin{equation}\begin{array}{lll}
\displaystyle E_n^{(0)+} = E_{n+1}^{(0)-},~~~~E_0^{(0)-} = 0\\
\displaystyle \psi_n^{+} = (E_{n+1}^{(0)-})^{-\frac{1}{2}}~A_0^+ \psi_{n+1}^{-}\\
\displaystyle \psi_{n+1}^{-} = (E_{n}^{(0)+})^{-\frac{1}{2}}~A_0^- \psi_{n}^{+},~ n= 0,1,2...
\end{array}
\end{equation}
Summing up, for a well behaved mass function the usual factorization technique for PDMSE allows us to construct nonsingular isospectral potentials of a given nonsingular potential using equations (\ref{e6}) and (\ref{e1}). It should be noted here that the factorization technique mentioned above is not unique and can be extended to use excited state wave functions. This generalization is mathematically straightforward but physically very nontrivial since it yields singular partner potentials. In the following we shall show that the excited state of a given position dependent mass Hamiltonian can be used to construct nonsingular isospectral potentials.

\subsection{Excited state factorization and its modification}
\noindent
In order to generalize the usual factorization technique to excited states we define the superpotential as
\begin{equation}
W_n(x) = - \frac{\psi_n^{-'}}{\sqrt{m}\psi_n^{-}}~~,n>0 \label{e2}
\end{equation}
where $\psi_n^{-}$ is the $n$-th excited state wave function of an exactly solvable nonsingular Hamiltonian $H_0^-$. It is clear that for a well behaved mass function the superpotential $W_n$ is singular function but for a suitably chosen singular mass function this becomes nonsingular. Since singular mass functions are physically absurd so we can not consider this type of mass function. Now subtracting the energy $E_n^{(0)-}$ of the excited state from the Hamiltonian so that the resulting Hamiltonian can be factorized, we have the form of the two partner potentials
\begin{equation}\begin{array}{ll}
\displaystyle V_n^-(x) =  W_n^2 - \left(\frac{W_n}{\sqrt{m}}\right)'  = V^-_0(x) -E_n^{(0)-}\\
\displaystyle V_n^+ (x) = V_n^-(x) + \frac{2 W'_n}{\sqrt{m}}- \frac{1}{\sqrt{m}}\left(\frac{1}{\sqrt{m}}\right)''
\end{array}\label{e5}
\end{equation}
The singular superpotential $W_n$ contributes nothing new to $V_n^-$ except a constant energy shift to $V^-_0$. In this case energy eigenvalues of the Hamiltonians $H^-_n$ and $H_0^-$ are related by $E_k^{(n)-} = E_k^{(0)-} - E_n^{(0)-}, k=0,1,2...$($k$ denotes the energy level and $n$ refers to the $n$th eigenfunction of $H_n^-$ used in factorization). But the partner potential $V_n^+$ becomes singular at the node(s) of the wave function $\psi_n^{-}$. This singularity is responsible for both the destruction of degeneracy of the spectrum and creation of negative energy state(s). The singularity breaks the real axis into more than one disjoint intervals and imposes additional boundary conditions on the wave functions. In this case, the supersymmetry operators map square integrable functions to states outside the Hilbert space. As a result, some or all of the wave functions of the SUSY partner Hamiltonians may not belong to the same Hilbert space of square integrable functions and usual proof of degeneracy between the excited states of the partner Hamiltonians does not hold \cite{JR84,Ca95}.

To overcome this, we require the modification of this excited state factorization technique which can be achieved by deforming the operators\footnote{$A_n^\pm$ are obtained from equation (\ref{e8}) after replacing $W_0$ by $W_n$.} $A_n^\pm$ as
\begin{equation}
\tilde{A}_n^\pm = \pm\frac{1}{\sqrt{m}}\frac{d}{dx} - \left(\frac{1 \mp 1}{2}\right) \left(\frac{1}{\sqrt{m}}\right)' + W_n + f_n
\end{equation}
where $W_n$ is given by equation (\ref{e2}) and the unknown function $f_n(x)$ will be evaluated shortly.
Using this operators $\tilde{A}_n^\pm$ we first obtain
\begin{equation}
\tilde{H}^+_n = \tilde{A}_n^+\tilde{A}_n^- = -\frac{1}{m}\frac{d^2}{dx^2} + \frac{m'}{m^2} \frac{d}{dx}+ \tilde{V}^+_n (x)
\end{equation}
where
\begin{equation}
\tilde{V}_n^+ = V_n^+ + \frac{f'_n}{\sqrt{m}} + \left(2 W_n + \frac{m' }{2 m^{\frac{3}{2}}}\right) f_n + f_n^2.
\end{equation}
At this point we assume
\begin{equation}
\frac{f'_n}{\sqrt{m}} + \left(2 W_n + \frac{m' }{2 m^{\frac{3}{2}}}\right) f_n + f_n^2 = \beta, ~~\beta \in \mathbb{R}\label{e3}
\end{equation}
 so that $\tilde{H}_n^+ = H_n^+ + \beta$ is solvable. This assumption also ensures the non uniqueness of the factorization viz.
  \begin{equation}
  A_n^+ A_n^- + \beta = \tilde{A}_n^+ \tilde{A}_n^-.
  \end{equation}
  For $\beta=0$, the Riccati equation (\ref{e3}) reduces to the Bernoulli equation whose solution is given by
\begin{equation}
f_n(x) = \frac{(\psi^{-}_n)^2}{\sqrt{m}\left(\lambda + \int (\psi^{-}_n)^2 dx\right)},~~~~\lambda \in \mathbb{R}.\label{e9}
\end{equation}
For nonzero $\beta$ the Riccati equation (\ref{e3}) is not always solvable but it can be transformed into the following second order linear differential equation
\begin{equation}
-\frac{1}{m}\psi^{+''} +\frac{m'}{m^2} \psi^{+'} + \left[W_n^2 + \left(\frac{W_n}{\sqrt{m}}\right)'+ \frac{W_n m'}{m^{\frac{3}{2}}}-\frac{1}{\sqrt{m}}\left(\frac{1}{\sqrt{m}}\right)''\right]\psi^{+} =-\beta \psi^+ \label{e4}
\end{equation}
with the help of following transformation
\begin{equation}
f_n(x) = \frac{1}{\sqrt{m}} (log \chi_n)',~~~~ \chi_n = \frac{e^{-\int \sqrt{m} W_n dx}}{\sqrt{m}} \psi^+.\label{e16}
\end{equation}
Using the relation (\ref{e5}) it is clear that the equation (\ref{e4}) is (nearly) isospectral to the following solvable equation
\begin{equation}
-\frac{1}{m}\psi^{-''} +\frac{m'}{m^2} \psi^{-'} + V_0^-(x) \psi^- = ( E_n^{(0)-}- \beta) \psi^-\label{e7}
\end{equation}
where the solutions $\psi^\pm (x)$ corresponding to two equations (\ref{e4}) and  (\ref{e7}) are related by
\begin{equation}
\psi^+ \sim A_n^+ \psi^-.\label{e17}
\end{equation}
By changing the order of operations between $\tilde{A}_n^+$ and $\tilde{A}_n^-$ and using relations (\ref{e5}), (\ref{e3}) we obtain the partner Hamiltonian $\tilde{H}_n^-$ of $\tilde{H}_n^+$ as
\begin{equation}
\tilde{H}_n^- = \tilde{A}_n^- \tilde{A}_n^+ = -\frac{1}{m} \frac{d^2}{dx^2} + \frac{m'}{m^2}\frac{d}{dx} + \tilde{V}^-_n(x)
\end{equation}
where
\begin{equation}
\tilde{V}_n^- = V_n^- - \frac{2 f_n'}{\sqrt{m}} + \beta = V_0^- - \frac{2 f_n'}{\sqrt{m}} - E_n^{(0)-} + \beta.\label{e14}
\end{equation}
At this point few comments are worth mentioning. For a given nonsingular potential $V_0^-$ and a mass function the new potential $\tilde{V}_n^+ (= V_n^+ +\beta)$ is singular but the potential $\tilde{V}_n^-$ has no singularity if $f_n(x)$ is nonsingular in which case the corresponding spectrum of the two isospectral Hamiltonians $\tilde{H}_n^-$ and $H_n^-$ are related by
\begin{equation}
\tilde{E}_k^{(n)-} = E_k^{(n)-} + \beta,~~~k=0,1,2....
\end{equation}
For $\beta = 0$, the singularity of $f_n$ given in (\ref{e9}) can be controlled by the arbitrary constant $\lambda$ and in all other cases it depends on the nature of solution of the equation (\ref{e7}).
 Moreover, the eigenfunctions of the two nonsingular Hamiltonians $H_n^-$ and $\tilde{H}_n^-$ are related by
\begin{equation}
\begin{array}{ll}
\displaystyle \tilde{\psi}^{-}_k~ \sim ~\tilde{A}_n^- A_n^+ \psi_k^{-}\\
\displaystyle \psi^{-}_k ~\sim ~ A_n^- \tilde{A}_n^+ \tilde{\psi}_k^{-}
\end{array}\label{e10}
\end{equation}
except for the zero energy state of $\tilde{H}_n^-$. The proof of this results is as follows: we start with
\begin{equation}\begin{array}{llll}
\tilde{H}_n^- (\tilde{A}_n^- A_n^+ \psi_k^-) = \tilde{A}_n^- (\tilde{A}_n^+ \tilde{A}_n^-) A_n^+ \psi_k^- \\
= \tilde{A}_n^- (A_n^+ {A}_n^- +\beta) A_n^+ \psi_k^- \\
= \tilde{A}_n^- A_n^+ ({A}_n^-A_n^+) \psi_k^- +\beta \tilde{A}_n^- A_n^+ \psi_k^- \\
= \tilde{A}_n^- A_n^+ (H_n^- \psi_k^-) +\beta \tilde{A}_n^- A_n^+ \psi_k^- = (E_k^{(n)-}+\beta)\tilde{A}_n^- A_n^+\psi_k^-.
\end{array}\label{e11}
\end{equation}
In the second step we have used the non-uniqueness of the factorization of $H_n^+$ and in the last step we have used the PDMSE $H_n^-\psi_k^- = E_k^{(n)-}\psi_k^-$. Hence, if $\psi_k^-$ is a eigenfunction of $H_n^-$ corresponding to eigenvalue $E_k^{(n)-}$ then $ \tilde{A}_n^- A_n^+\psi^-_k$ is a eigenfunction of $\tilde{H}_n^-$ with eigenvalue $E_k^{(n)-} + \beta.$ The second relation of equation (\ref{e10}) can be proved in a analogous way. The wave function $\tilde{\psi}_{k'}^-$ (say) corresponding to zero energy of the Hamiltonian $\tilde{H}_n^-$ can be obtained by directly solving $\tilde{A}_n^+ \tilde{\psi}^{-}_{k'} =0$.

Summing up, excited state wave function of a given position dependent mass Hamiltonian can be used to construct nonsingular partner Hamiltonian $\tilde{H}_n^-$ using equation (\ref{e14}) provided the function $f_n$ is nonsingular. This has been made possible in two step factorization. In the first step, an intermediate singular Hamiltonian $H_n^+$ has been created and in the second step non-uniqueness of the factorization technique has been used to remove the singularities. The operator $-\tilde{A}_n^- A_n^+$ is similar to the second order reducible intertwining operator ${\cal{L}} =\frac{1}{m(x)} \frac{d^2}{dx^2} + \eta(x) \frac{d}{dx} + \gamma(x)$ mentioned in ref.\cite{MR10} which intertwines two Hamiltonians $H_n^-$ and $\tilde{H}_n^-$
\begin{equation}
\tilde{H}_n^- \tilde{A}_n^- A_n^+ = \tilde{A}_n^- A_n^+ H_n^-,
\end{equation}
if one identifies $-\eta(x) = \frac{f_n}{\sqrt{m}} + \frac{m'}{m^2}$ and $-\gamma(x) = V_n^- + f_n W_n.$ In this regard the present algorithm is equivalent to the confluent second order SUSY transformation \cite{MN00} in constant mass case but the relation between the two isospectral partner potentials is different in the present approach due to the position dependence of the mass function. Moreover, the energy levels of the two nonsingular isospectral partner Hamiltonians $H_n^-$ and $\tilde{H}_n^-$ are degenerate. In figure 1, we have shown schematically the combined action of the operators $\tilde{A}_n^- A_n^+$ and $A_n^- \tilde{A}_n^+ $ on the wave function of two isospectral partner Hamiltonians $H_n^-$ and $\tilde{H}_n^-$ respectively. The operators $A_n^+$ or $\tilde{A}_n^+$ destroy a node but $\tilde{A}_n^-$ or $A_n^-$ create the same in the eigenfunctions so that the overall number of nodes remains the same.

\begin{figure}[h]
\begin{center}
\includegraphics[scale=.8]{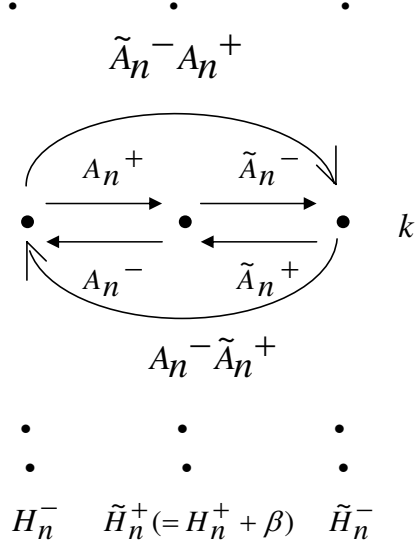}
\caption{\label{fig}Energy levels of two partner Hamiltonians $H_n^-$, $\tilde{H}_n^-$ and the intermediate Hamiltonian $\tilde{H}_n^+$. Combined actions of the operator $\tilde{A}_n^- A_n^+$ and $A_n^- \tilde{A}_n^+ $ are shown.}
\end{center}
\end{figure}
\section{Illustration}
It is worth mentioning here that the proposed technique mentioned in the preceding section is more general and can be applied to any solvable PDM Hamiltonian to obtain some acceptable isospectral Hamiltonians which might be useful in various fields of condensed matter physics. For illustration purpose we are considering here two exactly solvable position dependent mass Hamiltonians. In example 1 we have considered $\beta=0$ while non zero $\beta$ has been considered in example 2.

{\bf Example 1 ($\beta = 0$):~~} We consider the following mass function and potential possessing harmonic oscillator like spectra \cite{Pe08}:
\begin{equation}
m(x) = \frac{1}{1 + \alpha^2 x^2}
\end{equation}
\begin{equation}
V_0^- (x) = \left(\frac{\sinh^{-1} (\alpha x)}{\alpha}\right)^2 - \frac{\alpha^2}{4} \left(\frac{2 + \alpha^2 x^2}{1+ \alpha^2 x^2}\right),~~~ x\in(-\infty,\infty).
\end{equation}
The bound state solution are given by
\begin{equation}
\psi_k^{-} = \sqrt{\frac{1}{2^k k!}} \left(\frac{1}{\pi}\right)^{\frac{1}{4}} \frac{e^{-\frac{1}{2}~\left(\frac{\sinh^{-1} (\alpha x)}{\alpha}\right)^2}}{(1 + \alpha^2 x^2)^{1/4}}~ H_k\left(\frac{\sinh^{-1} (\alpha x)}{\alpha}\right)
\end{equation}
\begin{equation}
E_k^{(0)-} = 2 k +1, ~~~~k=0,1,2...
\end{equation}
where $H_k(x)$ denotes the Hermite polynomial. Now we consider here first excited state $\psi_1^{-} = \frac{2\sinh^{-1} x ~e^{-\frac{1}{2}\left(\sinh ^{-1} x\right)^2}}{(1 + x^2)^{1/4}}$, $\beta=0,$ and $\alpha=1$ for which
the partner potential $V_1^-$ and $\tilde{V}_1^-$ are obtained using equation (\ref{e5}), (\ref{e9}) and (\ref{e14}) as
\begin{equation}
V_1^- = \left({\sinh^{-1} ( x)}\right)^2 - \frac{1}{4} \left(\frac{2 +  x^2}{1+  x^2}\right) -3
\end{equation}
\begin{equation}
\tilde{V}_1^- = -\frac{2 + x^2}{4 + 4 x^2} + (\sinh^{-1} x )^2 -3 + 2 \sqrt{(1+x^2)} \frac{d}{dx} \left[\frac{4 \left(\sinh^{-1} x\right)^2}{\sqrt{\pi}e^{(\sinh^{-1} x)^2}(2 \lambda+ \ef(\sinh^{-1} x))-2 \sinh^{-1} x}\right]
\end{equation}
respectively. The energy eigenvalues of the Hamiltonian $\tilde{H}_1^-$ are given by $\tilde{E}^{(1)-}_k = 2k -2, ~k=0,1,2...$ The normalized first excited state wave function corresponding to zero energy of $\tilde{H}_1^-$ is evaluated by solving $\tilde{A}_1^+ \tilde{\psi}^{-}_1 = 0$ as
\begin{equation}
\tilde{\psi}_1^{-} = \frac{\pi^{\frac{1}{4}} \sqrt{8\lambda^2 -2}~~ e^{\frac{(\sinh^{-1} x)^2}{2}}\sinh^{-1} x }{(1+x^2)^{\frac{1}{4}}\left[2\sqrt{\pi}\lambda e^{(\sinh^{-1} x)^2} - 2 \sinh^{-1} x +\sqrt{\pi} e^{(\sinh^{-1} x)^2} \ef(\sinh^{-1} x)\right]}.
\end{equation}
Hence the new potential $\tilde{V}_1^{(1)-}$ is nonsingular for $\lambda>1/2$. The lowest energy wavefunction corresponding to negative energy $-2$ is obtained through $\tilde{A}_1^{-} A_1^+ \psi^{-}_0$ as
\begin{equation}
\tilde{\psi}^{-}_0 = \frac{\pi^{\frac{1}{4}}[2 \lambda + \ef(\sinh^{-1} x)] e^{\frac{(\sinh^{-1} x)^2}{2}}}{(1+x^2)^{\frac{1}{4}}\left[2\sqrt{\pi} \lambda e^{(\sinh^{-1} x)^2} - 2 \sinh^{-1} x + \sqrt{\pi} \ef(\sinh^{-1} x) e^{(\sinh^{-1} x)^2}\right]}.
\end{equation}
In figure 2(a), we have plotted the potential $V_1^-(x)$, its isospectral partner $\tilde{V}_1^-$ for two parameter values $\lambda=1, .7$ and the mass function. In figure 2(b), lowest and first excited state wave functions of the Hamiltonian $\tilde{H}_1^-$ for $\lambda=1$ have been plotted.
\begin{figure}[h]
\begin{center}
\includegraphics[scale=0.7]{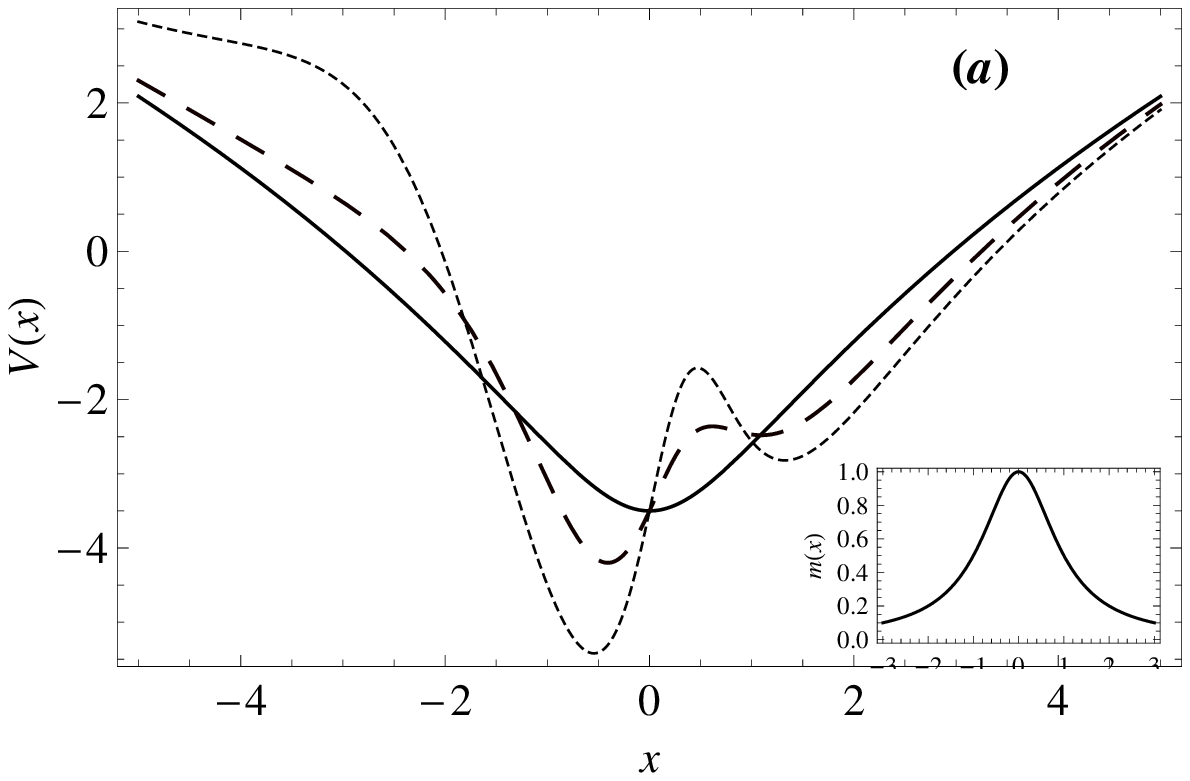}~~~~~~\includegraphics[scale=0.78]{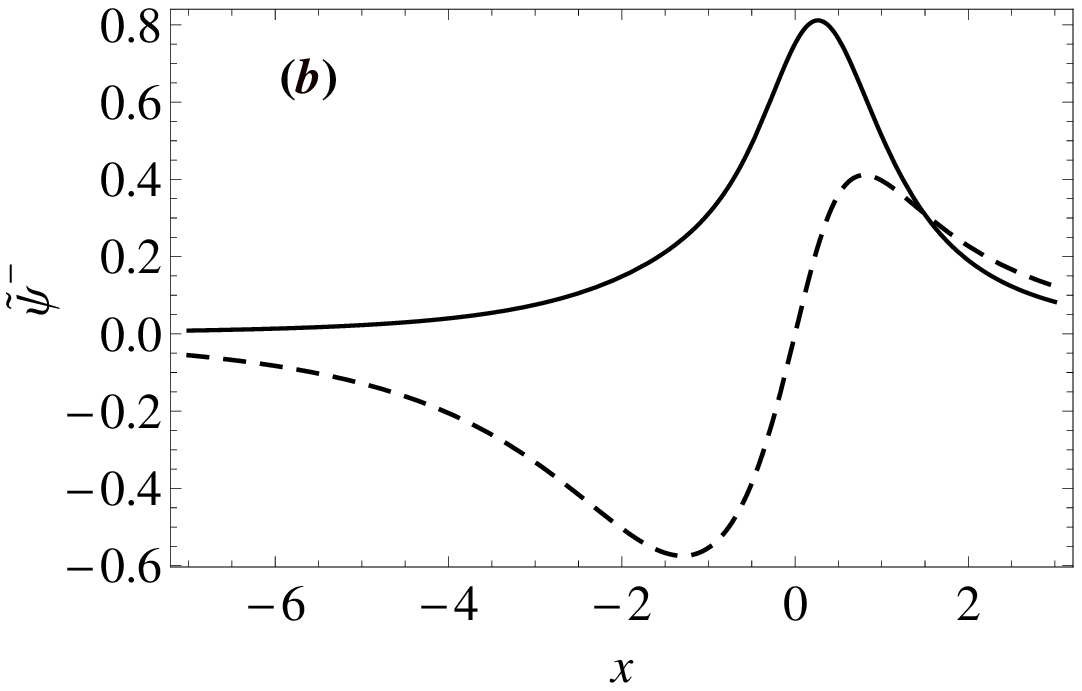}
\caption{\label{fig} (a) Plots of the given potentials $V_1^-$ (solid line) and $\tilde{V}_1^-$ for $\lambda =1$ (dashed line), $\lambda=.7$ (dotted line) and the mass function $m(x)$. (b) The lowest (solid line) and first excited state (dashed line) wave functions of the Hamiltonian $\tilde{H}_1^-$ for $\lambda =1.$}
\end{center}
\end{figure}

{\bf Example 2 ($\beta \ne 0$):~~} Here we consider the following mass function and the potential
\begin{equation}
m(x) = \frac{1}{4} sech^2 ~(\frac{x}{2})\label{e18}
\end{equation}
\begin{equation}
V_0(x) = \frac{(a+b-c)^2 - 1}{4} e^x + \frac{c(c-2)}{4} e^{-x},~~c>\frac{1}{2}, a+b-c+\frac{1}{2}>0,~x\in(-\infty,\infty),
\end{equation}
for which one linearly independent solution of the equation (\ref{e7}) is known to be (see appendix of ref.\cite{MR10} and \cite{Ma})
\begin{equation}
\psi^-(x) = C_1 ~e^{c x/2}(1+e^x)^{\frac{P-1}{2}}  ~~{_2}F_1\left(\frac{a+b+P}{2},\frac{-a-b+2c+P}{2},c,e^{-x}\right),\label{e19}
\end{equation}
where
\begin{equation}\begin{array}{ll}
P^2 = (a+b)^2 -2 c(a+b-c+1) + 4 (E_n^{(0)-}-\beta),\\
E_n^{(0)-} = n^2 + n(a+b) + \frac{c(a+b-c+1)}{2}.
\end{array}
\end{equation}
The bound state wave functions of the Hamiltonian $H_0^-$ are given \cite{MR10}, in terms of Jacobi polynomial $\mathcal{P}_n^{(\sigma,\delta)}(x)$, as
\begin{equation}
\psi_n^- \sim \frac{e^{cx/2}}{(1+e^x)^{(a+b+1)/2}}~~ \mathcal{P}_n^{(c-1,a+b-c)}\left(\frac{1-e^x}{1+e^x}\right),~~n=0,1,2...\label{e15}
\end{equation}
Now we consider first excited state i.e. $n=1$ for which the function $f_n$ can be obtained using equations (\ref{e16}) and (\ref{e17}) as
\begin{equation}
f_n= \frac{1}{\sqrt{m}}\frac{d}{dx} (\log[A_1^+ \psi^-(x)])+\frac{\cosh \left(\frac{x}{2}\right)}{4} \left(\frac{2 c}{(a+b-c+1)e^x -c} + \frac{a+b+3}{1+e^x}+c-a-b-1\right) + \sinh \left(\frac{x}{2}\right)
\end{equation}
where $A_1^+ = \frac{1}{\sqrt{m(x)}}\left(\frac{d}{dx}-\frac{\psi_1^{-'}}{\psi_1^-}\right)$,  $m(x), \psi^-$ and $\psi_1^-$ are given by equations (\ref{e18}), (\ref{e19}) and (\ref{e15}) respectively. In order to obtain new nonsingular potential $\tilde{V}_1^-$ we have to choose $a,b,c, \beta$ such that $f_n$ has no singularity. The analytical expression of $f_n$ is too involved so it is very difficult to find the range of parameters values for which $f_n$ is nonsingular. In figure 3(a), we have plotted the potential $V_1^-(x) = V_0^- - E_1^{(0)-}$ and its nonsingular partner $\tilde{V}_1^- = V_1^- -2f_n/\sqrt{m}-E_1^{(0)-}+\beta$ for particular values of $a=1,b=5, c=4$ and $\beta=1$. For the same set of parameter values, isospectral partner potential $\tilde{V}_2^-$ has been drawn in figure 3(b) by considering second excited state factorization i.e. $n=2$. The energy eigenvalues of the new Hamiltonians $\tilde{H}_1^-$ and $\tilde{H}_2^-$ are given by $\tilde{E}_k^{(1)-} = k^2+ 6k -6$ and $\tilde{E}_k^{(2)-} = k^2 + 6 k-15$ respectively. In both cases the wave functions can be obtained using equations (\ref{e10}) and (\ref{e15}).
\begin{figure}[h]
\begin{center}
\includegraphics[scale=.68]{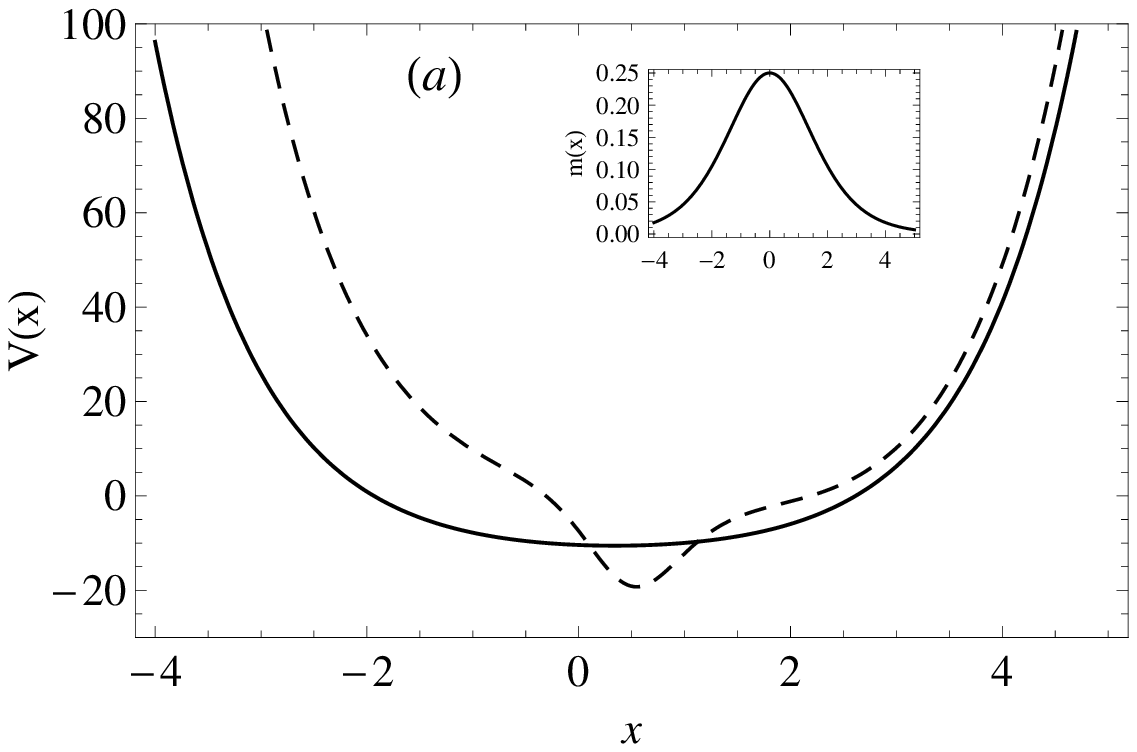}~~~\includegraphics[scale=.7]{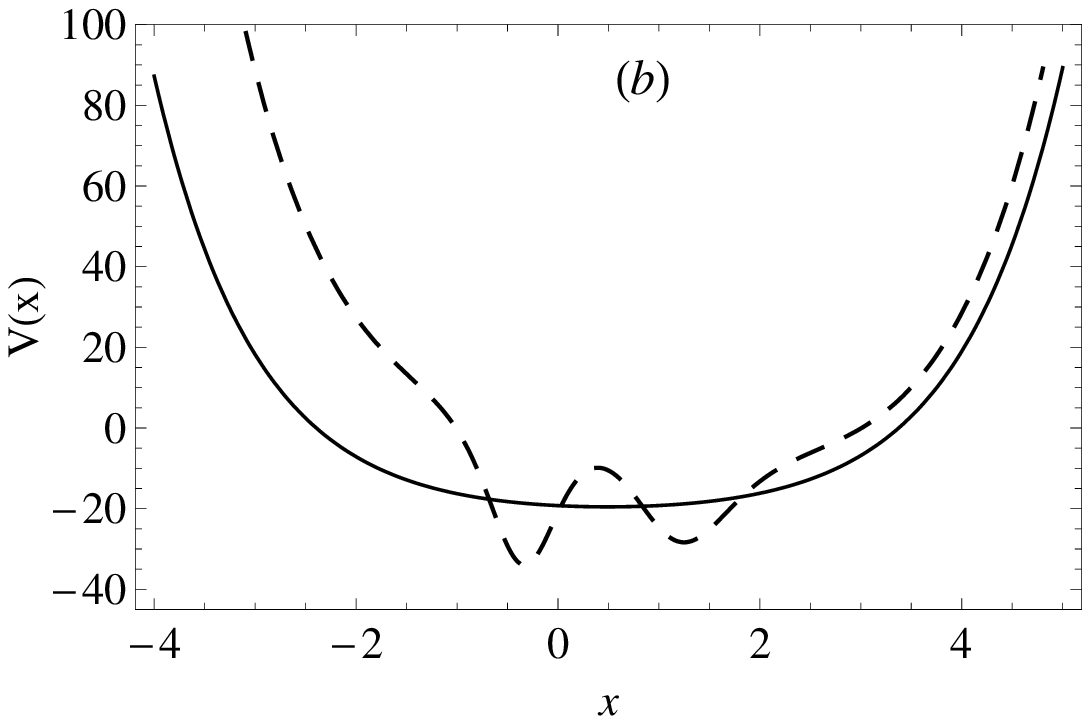}
\caption{\label{fig} Plots of (a) the potential $V_1^-=V_0^- - 13$ (solid line), its nonsingular partner $\tilde{V}_1^-$ (dashed line) and the mass function $m(x)$, (b) the potential $V_2^-=V_0^- -22$ (solid line) and its partner $\tilde{V}_2^-$ (dashed line) for $a=1, b=5, c=4, \beta=1$.}
\end{center}
\end{figure}
\section{Summary}
To conclude, we have generalized the modified factorization technique of a Schr\"{o}dinger Hamiltonian suggested by Berger and
Ussembayev \cite{BU10} to a quantum system characterized
by position-dependent mass Hamiltonian. This generalization is done in such a way that a given mass function and the excited state wave function of a PDM Hamiltonian can be used to generate new solvable physical Hamiltonians. The method discussed here is applied to a number of exactly solvable PDM Hamiltonians which leads to nontrivial isospectral Hamiltonians. In the constant mass limit of the present algorithm, all the recently published results \cite{BU10} on modified factorization can be recovered.

It would be interesting to study the modified factorization technique for a quantum system characterized by broken supersymmetry.


\begin{thebibliography}{100}
\bibitem{BD66} BenDaniel D J and Duke C B 1966 {\it Phys. Rev.} {\bf B 152} 683; Levy-Leblond J M 1995 {\it Phys. Rev.} {\bf A 52} 1845.
\bibitem{GK93} Geller M R and Kohn W 1993 {\it Phys. Rev. Lett.} {\bf 70} 3103; Serra L and Lipparini E 1997 {\it Europhys. Lett.} {\bf 40} 667; Arias de Saavedra F, Boronat J, Polls A and
Fabrocini A 1994 {\it Phys. Rev.} {\bf B 50}, 4248; Barranco M et al. 1997 {\it Phys. Rev.} {\bf B 56} 8997.
\bibitem{Pr65} Preston M A 1995 {\it Physics of the nucleus} (Addison-Wesley, INC, Reading); Luttinger J M and Kohn W 1955 {\it Phys. Rev.} {\bf 97} 869.
\bibitem{GG04}  Goser K, Glösekötter P and Dienstuhl J 2004 {\it Nanoelectronics and Nanosystems. From
Transistors to Molecular and Quantum Devices} (Springer, Berlin).
\bibitem{Ba88} Bastard G 1988 {\it Wave machanics Applied to Semiconductor
Heterostructure} (Les Editions de Physique, Les Ulis).
\bibitem{Ra01} Radovanovic J, Todorovic G, Milanovic V, Ikonic Z and Indjin D 2001 {\it Phys. Rev.} {\bf B 63} 115327.
\bibitem{Bo90} Bois P et.al 1990 {\it Superlatt. Microstruct.} {\bf 8}, 369; Tomic S, Milanovic V, Ikonic Z 1998 {\it Phys. Lett.} {\bf A 238}, 385.
\bibitem{Pl99} Plastino A R, Rigo  A, Casas M, Garcias F and Plastino A 1999 {\it Phys. Rev.} {\bf A 60} 4318.
 \bibitem{MI99} Milanovic V and Ikonic Z 1999 {\it J. Phys. A} {\bf 32} 7001.
 \bibitem{Go02}  Gonul B, Gonul B, Tutcu D and Ozer O 2002 {\it Mod. Phys. Lett.} {\bf A 17} 2057.
 \bibitem{GN07} Ganguly A and Nieto L M 2007 {\it J. Phys. A}  {\bf 40} 7265.
\bibitem{Al02} Alhaidari A D 2002 {\it Phys. Rev.} {\bf A 66} 042116.
 \bibitem{Ba05} Bagchi B, Gorain P, Quesne C and Roychoudhury R 2005 {\it Europhys. Lett.} {\bf 72} 155; Midya B and Roy B 2009 {\it Phys. Lett.} {\bf A 373} 4117.
 \bibitem{Qu09} Quesne C 2009 {\it SIGMA} {\bf 5} 046; Ganguly A, Ioffe M V and Nieto L M 2006 {\it J. Phys. A } {\bf 39} 14659.
\bibitem{RR02} Kock R and Koca M 2003 {\it J. Phys. A} {\bf 36} 8105; Roy B and Roy P 2002 {\it J. Phys. A} {\bf 35} 3961.
\bibitem{DC99} Dekar L, Chetouani L and Hammann T F 1999 {\it Phys. Rev.} {\bf A 59} 107.
\bibitem{MR10} Midya B, Roy B and Roychoudhury R 2010 {\it J. Math. Phys.} {\bf 51} 022109.
\bibitem{SH08} Tanaka T 2006 {\it J. Phys. A} {\bf 39} 219.
\bibitem{IH51} Infeld L and Hull T E 1951 {\it Rev. Mod. Phys.} {\bf 23} 21; Mielnikh B 1984 {\it J. Math. Phys.} {\bf 25} 3387; Sukumar C V 1985 {\it J. Phys. A} {\bf 18} 2917.
\bibitem{An84} Andrianov A A, Borisov T E and Ioffe M V 1984 {\it Phys. Lett.} {\bf A 105} 19; Andrianov A A, Ioffe M V and Spiridonov V 1993 {\it Phys Lett} {\bf A 174} 273; Andrianov A A et al 1995 {\it Int. J. Mod. Phys.} {\bf A 10} 2683.
 \bibitem{FG04} Fernandez D J and Garcia N F 2004 {\it AIP Conf. Proc.} {\bf 744} 236; Fernandez D J 1997 {\it Int. J. Mod. Phys.} {\bf A 12} 171; Mielnik B and Rosas-Ortiz O 2004 {\it J. Phys. A} {\bf 37} 10007. 
\bibitem{BU10}  Berger M S and Ussembayev N S 2010 {\it Phys. Rev.} {\bf A 82} 022121;  Berger M S and Ussembayev N S 2010 {\it J. Phys. A} {\bf 43} 385309; Datta D and  Roy P 2011 {\it Phys. Rev.} {\bf A 83} 054102.
\bibitem{JR84} Jevicki A and Rodrigues J P 1984 {\it Phys. Lett.} {\bf B 146} 55; Panigrahi P K and Sukhatme U P 1993 {\it Phys. Lett.} {\bf A 178} 1993.
\bibitem{Ca95} Casahorran J 1995 {\it Physica} {\bf A 217} 429; Robnik M 1997 {\it J. Phys. A} {\bf 30} 1287.
\bibitem{MN00} Mielnik B, Nieto L M and Rosas-Ortiz O 2000 {\it Phys. Lett.} {\bf A 269} 70; Fernandez D J and Salinas-Hernandez E 2003 {\it J. Phys. A } {\bf 36} 2537; Fernandez D J, Salinas-Hernandez 2005 {\it Phys. Lett.} {\bf A338} 13.
\bibitem{Pe08} Pena J J et al. 2008 {\it Int. J. Quant. Chem.} {\bf 108} 2906.
\bibitem{Ma} This solution can also be obtained using Mathematica.
\end{thebibliography}
\end{document}